\documentclass[10pt,twocolumn,twoside]{IEEEtran}

\usepackage{tikz}
\usepackage{amsmath,amsfonts,amssymb}%
\usepackage{bm}%
\usepackage{graphicx,graphics}%
\usepackage{cases}%
\usepackage[noadjust]{cite}%
\usepackage{color}%
\usepackage{cite,url}
\usepackage{verbatim}
\usepackage{algorithm}
\usepackage{balance}
\usepackage{multirow}
\usepackage{stfloats}

\usepackage{xspace}
\usepackage{mathtools} 
\usepackage{subfig} 

\usepackage{caption}

\usepackage{tikz}
\usetikzlibrary{shapes.misc}
\usetikzlibrary{matrix}
\usetikzlibrary{arrows,backgrounds,fit,calc}
\usepackage{algorithm}
\usepackage{algpseudocode}
\usepackage{ifpdf}

\ifpdf
  \usepackage[pdftex,colorlinks=true,breaklinks=true,citecolor=blue]{hyperref}
\else
  \usepackage[ps2pdf,colorlinks=true,citecolor=blue]{hyperref}
\fi

\ifCLASSOPTIONonecolumn
    
    \usepackage[margin=25mm]{geometry}
\fi

\usepackage{float}
\floatname{algorithm}{Procedure}

\newcommand{\secref}[1]{Section~\ref{#1}}

\DeclarePairedDelimiterX\abs[1]{\lvert}{\rvert}{#1}
\DeclarePairedDelimiterX\parn[1]{(}{)}{#1}
\DeclarePairedDelimiterX\set[1]{\lbrace}{\rbrace}{#1}
\DeclarePairedDelimiterX\innerp[2]{\langle}{\rangle}{#1,#2}
\DeclarePairedDelimiterX\norm[1]{\lVert}{\rVert}{#1}
\DeclarePairedDelimiterX\brak[1]{\lbrace}{\rbrace}{#1}
\DeclarePairedDelimiterX\coeff[1]{(}{)}{#1}

\newcommand{\sot}{\mathrm{SO(3)}}
\newcommand{\lsot}{L^{2}(\sot)}
\newcommand{\sohc}[4]{\big({#1} \big){}^{#2}_{#3,#4}}

\newcommand{\conj}[1]{\overline{#1}} 

\newcommand{\dfn}{\triangleq}

\newcommand{\figref}[1]{Fig.\,\ref{#1}}

\newtheorem{remark}{Remark}

\graphicspath{{figs/},{Figures/}}

\newcommand{\lsphL}[1]{\mathcal{H}_{#1}}
\newcommand{\ssk}[1]{\mathfrak{K}_{L}}
\newcommand{\ssgl}[1]{\mathfrak{G}_{L}}
\newcommand{\sso}[1]{\mathfrak{E}_{L}}

\newcommand{\nks}{N(\ssk{L})}
\newcommand{\ngls}{N(\ssgl{L})}

\newcommand{\matlab}{\texttt{MATLAB}}

\begin{document}
\title{Gauss-Legendre Sampling on the Rotation Group}%
\author{Zubair Khalid, Salman Durrani, Rodney A. Kennedy, Yves Wiaux and Jason D. McEwen

\thanks{Z.~Khalid is with the School of Science and Engineering, Lahore University of Management Sciences, Lahore 54792, Pakistan. S.~Durrani and R.~A.~Kennedy are with
   the Research School of Engineering, College of Engineering and
   Computer Science, Australian National University, Canberra, ACT 2601,
   Australia. Y.~Wiaux is with the Institute of Sensors, Signals \& Systems, Heriot-Watt University, Edinburgh EH14 4AS, UK.
    J.~D.~McEwen is with the Mullard Space Science Laboratory, University College London, Surrey RH5 6NT, UK. S.~Durrani, R.~A.~Kennedy and J.~D.~McEwen are partially supported by the Australian Research Council's Discovery Projects funding scheme (Project no.~DP150101011). J.~D.~McEwen is also supported by the Engineering and Physical Sciences Research Council (Grant no.~EP/M011852/1).}
\thanks{E-mail: zubair.khalid@lums.edu.pk, salman.durrani@anu.edu.au, rodney.kennedy@anu.edu.au, y.wiaux@hw.ac.uk, jason.mcewen@ucl.ac.uk
}
}
\maketitle

\begin{abstract}
We propose a Gauss-Legendre quadrature based sampling on the rotation group for the representation of a band-limited signal such that the Fourier transform (FT) of a signal can be exactly computed from its samples. Our figure of merit is the sampling efficiency, which is defined as a ratio of the degrees of freedom required to represent a band-limited signal in harmonic domain to the number of samples required to accurately compute the FT. The proposed sampling scheme is asymptotically as efficient as the most efficient scheme developed very recently. For the computation of FT and inverse FT, we also develop fast algorithms of complexity similar to the complexity attained by the fast algorithms for the existing sampling schemes. The developed algorithms are stable, accurate and do not have any pre-computation requirements. We also analyse the computation time and numerical accuracy of the proposed algorithms and show, through numerical experiments, that the proposed Fourier transforms are accurate with errors on the order of numerical precision.
\end{abstract}
\vspace{-1mm}
\begin{IEEEkeywords}
rotation group, $\sot$, sampling, band-limited signals, Fourier transform.
\end{IEEEkeywords}
\vspace{-1mm}
\vspace{-1mm}

\section{Introduction}

The processing and analysis of signals defined on the rotation group, denoted by $\sot$, in the harmonic domain is crucial in applications found in various fields of science and engineering (e.g., \cite{Kazhdan:2003,Schaeben:2003,Hoover:2008,Khalid:2013,Kavocs:2003,Chirikjian:2000,mcewen:2006:fcswt}). The transformation of a signal on $\sot$ to the harmonic domain is enabled by the Fourier transform (FT) on the rotation group, which is, sometimes, also termed as Wigner-$D$ transform in the literature~\cite{Kostelec:2008,Potts:2009}. Clearly, the ability to compute the FT on the rotation group from a finite number of samples of the signal is of significant importance. In this work, we focus on the reduction in the number of samples required for the representation of a band-limited signal defined on $\sot$ such that the FT can be exactly and efficiently (in terms of computational effort) determined from the samples of the signal.

The design of sampling schemes on the rotation group and the development of computationally efficient FT's on the rotation group have been actively investigated in the literature~\cite{Risbo:1996,Yershova:2004,Kostelec:2008,Potts:2009,Graf:2009,Hielscher:2010,Keiner:2012,Grafthesis2013}. Among the developments in the literature, we focus on the sampling schemes that form regular or equiangular grid of samples on the rotation group and support accurate computation of the FT of the signal band-limited to degree $L$~(formally defined in \secref{Sec:FT-preliminaries})~\cite{Potts:2009,Kostelec:2008,Risbo:1996}.

An exact method for the computation of the FT, based on a sampling theorem, has been proposed for an equiangular sampling scheme composed of $8L^3$ samples with $2L$ samples along each of the three Euler angles of $\sot$, \cite{Kostelec:2008}. Using the separation of variables, they also developed fast algorithm of complexity $O(L^4)$ for the computation of FT. They also point out that the algorithm of complexity $O(L^3 \log^2 L)$ can also be developed as a variant of the algorithm in \cite{Driscoll:1994}. However, the algorithm of complexity $O(L^3 \log^2 L)$ requires pre-computation of $O(L^4)$ and is therefore only practical for smaller band-limits~($L < 256$) due to the storage requirements~\cite{Kostelec:2008}.

By expanding the Wigner-$d$ functions into Chebyshev polynomials and using the fast polynomial transform~\cite{Kunis:2003}, an algorithm for the computation of the inverse FT of complexity $O(L^4)$ has been developed. The inverse FT algorithm with reduced complexity $O(L^3 \log^2 L)$  has also been developed, which is shown to suffer from stability issues and therefore requires additional stabilization steps~\cite{Potts:1998,Suda:2002,Healy:2003}, making the exact complexity of the algorithm difficult to determine. For non-equispaced samples, the FT as an adjoint of the IFT has also been developed~\cite{Potts:2009}. However, it is applied, using the Clenshaw-Curtis quadrature rule, to the equiangular sampling scheme that consists of $8L^3$ number of samples.

The sampling efficiency, defined as a ratio of number of Fourier coefficients required to represent a band-limited signal in the Fourier domain to the number of samples required to accurately compute FT, is the fundamental property of any sampling scheme. In addition to the sampling efficiency, the computational complexity and accuracy of the FT associated with the sampling scheme are of significant importance. The existing sampling configurations that support exact (or sufficiently accurate) FT attain the sampling efficiency of one third as we show later in the paper. Furthermore, the algorithms associated with the existing sampling schemes, which are stable and accurately compute the FT and do not have any pre-computation requirements, have the computational complexity $O(L^4)$.

We summarise the contributions of this work as follows.

\begin{itemize}
\item We propose a Gauss-Legendre~(GL) quadrature based sampling scheme on the rotation group for the exact computation of the FT. The proposed sampling scheme is asymptotically as efficient as the equally most efficient sampling scheme developed very recently.\footnote{As an alternative to the proposed Gauss-Legendre quadrature based scheme, an equiangular sampling scheme, based on a sampling theorem, on the rotation group has been \emph{recently} proposed~\cite{Jason:so3:2015} with the same asymptotic sampling efficiency that is attained by the proposed sampling scheme.}

\item For the proposed sampling scheme, we develop fast algorithms of complexity $O(L^4)$ for the computation of the FT and IFT. The developed algorithms are stable, accurate and do not require any pre-computation.

\item We show, through numerical experiments, that FT can be computed to near machine precision accuracy.
\end{itemize}

We organize the rest of the paper as follows. The mathematical preliminaries for the signals on the rotation group are summarised in \secref{sec:preliminaries}. In \secref{sec:FT-and-schemes}, we formulate the FT on the rotation group, present the proposed sampling scheme and develop fast algorithm for the computation of Fourier transform. The computation time and the numerical accuracy are analysed in \secref{sec:analysis}. Finally, the conclusions are made in \secref{sec:conclusions}.

\section{Preliminaries}\label{sec:preliminaries}

\subsection{Signals on the Rotation Group}

We consider complex-valued 
functions $f(\varphi,\vartheta,\omega)$, defined on a rotation group $\sot$ parameterized by Euler angles $(\varphi,\vartheta,\omega)$, where $\varphi\in
[0,\,2\pi)$, $\vartheta\in [0,\,\pi]$ and $\omega \in
[0,\,2\pi)$. The set of square integrable complex-valued functions defined on $\sot$ forms a Hilbert
space $\lsot$ with the inner product for two functions $f$ and $h$ defined on $\sot$:
\begin{align}\label{eqn:innprd}
\langle f, h \rangle \triangleq
\int_{\sot} f(\varphi,\vartheta,\omega) \, \conj{h(\varphi,\vartheta,\omega)} \, d\varphi \, \sin\vartheta\,d\vartheta\, d\omega,
\end{align}
where $\overline{(\cdot)}$ denotes the complex conjugate
and the integral is a triple integral over all rotations $(\varphi,\vartheta,\omega)$~\cite{Kennedy-book:2013,Kostelec:2008,Graf:2009,Potts:2009}. The inner product in \eqref{eqn:innprd} induces a norm
$\|f\| \triangleq\langle f,f \rangle^{1/2}$. The functions
with finite induced norm are referred to as signals on the rotation group or signals for short.

\subsection{Wigner-$D$ Functions}

The Wigner-$D$ function, denoted by $D_{m,n}^{\ell}(\varphi,\vartheta,\omega)$, is defined for
degree $\ell$ and orders $|m|,|n|\leq\ell$ as~\cite{Potts:2009,Kostelec:2008,Kennedy-book:2013}
\begin{equation}\label{Eq:Dlm}
D^{\ell}_{m,n}(\varphi,\vartheta,\omega) = e^{-im\varphi}d_{m,n}^{\ell}(\vartheta)\,
e^{-in\omega},
\end{equation}
where $d_{m,n}^{\ell}(\vartheta)$ is the Wigner-$d$
function~\cite{Kennedy-book:2013}. We use the following decomposition of Wigner-$d$ function $d_{m,n}^{\ell}(\vartheta)$ in terms of complex exponentials~\cite{Risbo:1996}
\begin{align}\label{Eq:Wignerd-risbo}
d^{\ell}_{m,n}(\vartheta) = i^{n-m} \sum_{m'=-\ell}^{\ell} \Delta_{m',m}^{\ell} \Delta_{m',n}^{\ell} e^{im'\vartheta},
\end{align}
where $\Delta_{m,n}^{\ell} = d^{\ell}_{m,n}({\pi}/{2})$. Using the symmetry relations exhibited by Wigner-$d$ functions~\cite{Kennedy-book:2013}, the Wigner-$d$ function in \eqref{Eq:Wignerd-risbo} can be expressed as
\begin{align}\label{Eq:Wignerd-risbo-polynomial}
d^{\ell}_{m,n}(\vartheta) &= i^{n-m}\sum_{m'=0}^{\ell} \Delta_{m',m}^{\ell} \Delta_{m',n}^{\ell} X(m-n,m',\vartheta)
\end{align}
where
\begin{equation*}
X(m-n,m',\vartheta) =
\begin{cases}
1, & m'=0,\\
2\cos(m'\vartheta), & m'\ne0,\, m-n~{\rm even}, \\
 2i\,\sin(m'\vartheta), & m'\ne0,\, m-n~{\rm odd}.
\end{cases}
\end{equation*}

\begin{remark}[On the polynomial degree of Wigner-$d$ functions]
\label{remark:degree-Wigner}
Since $\cos(m'\vartheta) = T_{m'}(\cos\vartheta)$ and $\sin(m'\vartheta) = \sin\vartheta\, U_{m'-1}(\cos\vartheta)$, where $T_{m'}(\cos\vartheta)$ and $U_{m'}(\cos\vartheta)$ represent the Chebyshev polynomials of degree $m'$ of the first and second kind, respectively, it can be inferred from \eqref{Eq:Wignerd-risbo-polynomial} that $d^{\ell}_{m,n}(\vartheta)$ is a polynomial of degree $\ell$ in $\cos\vartheta$.
\end{remark}

\subsection{Fourier Transform~(FT) on the Rotation Group}
\label{Sec:FT-preliminaries}
The Wigner-$D$ functions for integer degree $\ell \geq0$ and integer orders $|m|,|n|\leq\ell$
form a complete set of orthogonal functions for the
space $\lsot$. The orthogonality relation for Wigner-$D$ functions is given by
\begin{align}\label{Eq:WignerD_ortho}
\innerp[\big]{ D^{\ell}_{m,n}}{D^{\ell'}_{m',n'}} =
 \frac{8 \pi^2}{2\ell +1}\,\delta_{\ell, \ell'} \delta_{m,m'}
\delta_{n,n'},
\end{align}
where $\delta_{\ell, \ell'}$ is the Kronecker delta. By completeness of the Wigner-$D$ functions, any function $f\in\lsot$ can be expressed as
\begin{align}\label{Eq:f_synthesis}
f(\varphi,\vartheta,\omega) =  \sum_{\ell=0}^\infty
\sum_{m=-\ell}^{\ell}\sum_{n=-\ell}^{\ell} \sohc{f}{\ell}{m}{n}
D^{\ell}_{m,n}(\varphi,\vartheta,\omega),
\end{align}
where
\begin{align}\label{Eq:f_analysis}
\sohc{f}{\ell}{m}{n} = \frac{2\ell +1}{8 \pi^2}\,
\innerp[\big]{f}{D^{\ell}_{m,n}},
\end{align}
denotes the \emph{Fourier coefficient} of degree $\ell$, orders $m$ and $n$ and forms the Fourier~(or spectral) domain representation
for signals defined on $\sot$.

The signal $f$ is said to be \emph{band-limited at degree $L$}
if $\sohc{f}{\ell}{m}{n}=0$, $\forall \ell\ge L$. The set of signals band-limited at degree $L$ form a subspace of $\lsot$, which is denoted by $\lsphL{L}$. Let
\begin{equation}
\label{Eq:dim_subspace}
d_L = \sum_{\ell=0}^{L-1}
\sum_{m=-\ell}^{\ell}\sum_{n=-\ell}^{\ell} 1 = \frac{L(2L-1)(2L+1)}{3}
\end{equation}
denote the dimension of the subspace $\lsphL{L}$.  Then $d_L$ also quantifies the degrees of freedom required in the Fourier domain to represent a signal band-limited at $L$. We refer to the computation of Fourier coefficients $\sohc{f}{\ell}{m}{n}$ from the signal $f$, given in \eqref{Eq:f_analysis}, as the \emph{Fourier transform}~(FT) of $f$. The synthesis of a signal $f$ from its Fourier coefficients $\sohc{f}{\ell}{m}{n}$, given in \eqref{Eq:f_synthesis}, is referred as \emph{inverse Fourier transform}~(IFT).

\subsection{Sampling on the Rotation Group}

In order to accurately determine the FT of a band-limited signal on the rotation group from its samples, an equiangular sampling scheme, denoted by $\ssk{L}$, has been proposed in \cite{Kostelec:2008}, which is composed of $8L^3$ samples on a regular grid formed by $2L$ number of samples along each Euler angle in $(\varphi,\vartheta,\omega)$.

We define the asymptotic sampling efficiency, denoted by $E(\cdot)$, of any sampling scheme as a ratio of the dimension $d_L$ of the subspace $\lsphL{L}$ formed by the band-limited signals to the number of samples, denoted by $N$, required to compute FT of a band-limited signal $f\in\lsphL{L}$ as $L\rightarrow \infty$. Since the sampling scheme $\ssk{L}$ requires $\nks = 8L^3$ number of samples, it attains an asymptotic sampling efficiency
\begin{equation}
E(\ssk{L}) = \lim_{L\rightarrow \infty} \frac{d_L}{\nks} = \frac{1}{6}.
\end{equation}
We note that the sampling scheme that also requires $8L^3$  number of samples has also been proposed in \cite{Potts:2009} for the computation of FT. By applying the periodic extension approach along $\vartheta$, originally proposed in \cite{McEwen:2011}, the sampling scheme with asymptotic efficiency $2\,E(\ssk{L})$ has been proposed just recently in~\cite{Jason:so3:2015}. Furthermore, the algorithms developed for the computation of FT and IFT for these existing schemes have the computational complexity $O(L^4)$.

\section{Proposed Sampling and Fourier Transform}
\label{sec:FT-and-schemes}

We propose a sampling scheme with an efficiency of $2E(\ssk{L})$ for the representation of band-limited signals defined on the rotation group. We formulate the FT and IFT for the proposed scheme and develop fast algorithms for their computation.

\subsection{Harmonic Formulation}

We first develop the formulation for FT and IFT.
For a band-limited signal $f\in\lsphL{L}$, we rewrite the formulation of IFT in \eqref{Eq:f_synthesis}, by changing the order of summation
and using \eqref{Eq:Dlm}, as
\begin{equation}\label{Eq:form-synthesis}
f(\varphi,\vartheta,\omega) = ~
\smashoperator[l]{\sum_{m=-(L-1)}^{L-1}}\;\sum_{n=-(L-1)}^{L-1}e^{-i(n\omega + m\varphi)} A_{m,n}(\vartheta),
\end{equation}
where
\begin{equation}\label{Eq:Amn}
A_{m,n}(\vartheta) = \sum_{\ell=0}^{L-1} \sohc{f}{\ell}{m}{n} d^{\ell}_{m,n}(\vartheta).
\end{equation}
Since $\sohc{f}{\ell}{m}{n}$, $d^{\ell}_{m,n}(\vartheta)$ and $\Delta_{m,n}^{\ell}$ are only defined for orders $|m|,|n|\leq\ell$, we define $\sohc{f}{\ell}{m}{n}=d^{\ell}_{m,n}(\vartheta) = \Delta_{m,n}^{\ell}=0$ for $|m|,|n|>\ell$ to facilitate the interchange of summations in the formulation of FT or IFT.

Using the formulation of IFT in \eqref{Eq:form-synthesis} and the orthogonality of complex exponentials, we
write the FT, \eqref{Eq:f_analysis}, as
\begin{align}\label{Eq:form-analysis}
\sohc{f}{\ell}{m}{n} &= \frac{2\ell +1}{2}\int_{0}^{\pi} A_{m,n}(\vartheta)\,d^{\ell}_{m,n}(\vartheta)\,\sin\vartheta\,d\vartheta,
\end{align}
where $A_{m,n}(\vartheta)$ can be recovered from $f(\varphi,\vartheta,\omega)$, again using \eqref{Eq:form-synthesis} and noting the orthogonality of complex exponentials, as
\begin{equation}
\label{Eq:Amn-analysis}
A_{m,n}(\vartheta) =  \frac{1}{4\pi^2}\int_{0}^{2\pi}  \int_{0}^{2\pi} f(\varphi,\vartheta,\omega)e^{i(n\omega + m\varphi)}\, d\omega \,d\varphi.
\end{equation}

\subsection{Gauss-Legendre Sampling}
The computation of the FT requires the evaluation of integrals over $\varphi$ and $\omega$ in \eqref{Eq:Amn-analysis} and an integral over $\vartheta$ in \eqref{Eq:form-analysis}. Since $\max(|m|)<L$ and $\max(|n|)<L$, the integrals involved in the computation of $A_{m,n}(\vartheta)$ can be replaced with summations if \emph{\emph{at least}} $2L-1$ samples of the signal are taken along each $\varphi$ and $\omega$. Once $A_{m,n}(\vartheta)$ is computed, the Gauss-Legendre quadrature rule can be used to discretise the integral over $\vartheta$ by taking only $L$ samples along $\vartheta$, instead of $2L$ samples considered in \cite{Potts:2009,Kostelec:2008}.

With these considerations, we propose a sampling scheme, denoted by $\ssgl{L}$, that is composed of $2L-1$ equiangular sample points along each $\varphi$ and $\omega$, given by
\begin{equation}
\label{Eq:equiangular_2dim}
\varphi_u = \omega_u = \frac{2\pi u}{2L-1},\quad u\in\{0,1,\dotsc,2L-2\}.
\end{equation}
and $L$ samples along $\vartheta$, denoted by $\vartheta_v$ for integer $v\in\{0,1,\dotsc,L-1\}$, 
as the roots of the Wigner-$d$ function
\begin{equation}
\label{Eq:Wignerd-roots}
d^{L}_{0,0}(\vartheta) = P_L(\cos\vartheta)=  0,
\end{equation}
where $P_L$ denotes the Legendre polynomial of degree $L$~\cite{Kennedy-book:2013}. Note that the sample points along $\vartheta$ as roots of the Wigner-$d$ function~(or Legendre polynomial) can be determined precisely~\cite{Press:1993}. We refer to the proposed scheme $\ssgl{L}$ as the \emph{Gauss-Legendre~(GL) sampling} scheme.

\subsection{Fourier Transform --- Computation}

The FT of band-limited signal $f\in\lsphL{L}$ sampled over $\ssgl{L}$ can be determined by first computing $A_{m,n}(\vartheta_v)$ for each $\vartheta_v$, following \eqref{Eq:Amn-analysis}, as
\ifCLASSOPTIONonecolumn
\begin{equation}
\label{Eq:Amn_discrete}
A_{m,n}(\vartheta_v) = \frac{1}{(2L-1)^2}
\sum_{u=0}^{2L-1}\sum_{w=0}^{2L-1} f(\varphi_u,\vartheta_v,\omega_w) e^{i(n\omega_w + m\varphi_u)},
\end{equation}
\else
\begin{multline}
\label{Eq:Amn_discrete}
A_{m,n}(\vartheta_v) = \frac{1}{(2L-1)^2} \\
\times \sum_{u=0}^{2L-1}\sum_{w=0}^{2L-1} f(\varphi_u,\vartheta_v,\omega_w) e^{i(n\omega_w + m\varphi_u)},
\end{multline}
\fi
where the equivalence between \eqref{Eq:Amn-analysis} and \eqref{Eq:Amn_discrete} can be readily verified by substituting \eqref{Eq:form-synthesis} in \eqref{Eq:Amn-analysis} or \eqref{Eq:Amn_discrete} and employing the orthogonality of continuous or discrete complex exponentials.

Once $A_{m,n}(\vartheta_v)$ in \eqref{Eq:Amn_discrete} is computed, the Fourier coefficient in $\sohc{f}{\ell}{m}{n}$ can be determined by discretising the integral in \eqref{Eq:form-analysis} using the GL quadrature rule\footnote{GL quadrature requires $L$ sampling points to compute the integral \emph{exactly} of the polynomial of degree $2L-1$.}. Since we have taken $L$ sample points along $\vartheta$ as roots of \eqref{Eq:Wignerd-roots}, we need to show that the integrand in \eqref{Eq:form-analysis} is a polynomial in $\cos\theta$ of degree at most $2L-1$, so that the GL quadrature can be applied to discretise the integral. Following Remark~\ref{remark:degree-Wigner} and the formulation for $A_{m,n}(\vartheta)$, we note that $A_{m,n}(\vartheta)$, given in \eqref{Eq:Amn}, is a polynomial of degree $L-1$. Similarly, the integrand $A_{m,n}(\vartheta) d^{\ell}_{m,n}(\vartheta)$ in \eqref{Eq:form-analysis} is a polynomial of degree $L-\ell-1$, which reaches a maximum of $2L-2$ for $\ell=L-1$. Therefore, the GL quadrature leads to
\begin{equation}\label{Eq:GL-quadrature-fomrulation}
\sohc{f}{\ell}{m}{n} = \frac{2\ell +1}{2} \sum_{v=0}^{L-1} A_{m,n}(\vartheta_v) d^{\ell}_{m,n}(\vartheta_v) q(\vartheta_v),
\end{equation}
where $q(\vartheta_v)$ denotes the GL quadrature weight assigned to each sample point $\vartheta_v$ and is given by
\begin{align}
q(\vartheta_v) = 2\Big(\frac{\sin\vartheta_v}{L\,\, d^{L-1}_{0,0}(\vartheta_v)} \Big)^2.
\end{align}

\begin{remark}[On the efficiency of GL sampling scheme]
The proposed GL scheme requires $\ngls \dfn L(2L-1)^2$ number of samples in total and thus attains an asymptotic sampling efficiency
\begin{equation}
E(\ssgl{L}) = \lim_{L\rightarrow \infty} \frac{d_L}{\ngls} = \frac{1}{3} = 2 E(\ssk{L}).
\end{equation}
Thus, the proposed GL sampling scheme is as efficient (asymptotically) as the most efficient scheme developed just recently~\cite{Jason:so3:2015}, and is more efficient, by a factor of two, than the schemes presented in \cite{Kostelec:2008} and \cite{Potts:2009}.
\end{remark}

\subsection{Fourier Transform --- Fast Algorithm}

Here we develop fast algorithm for the computation of FT, given in \eqref{Eq:GL-quadrature-fomrulation}, of a band-limited signal $f\in\lsphL{L}$ discretised over the proposed GL sampling scheme $\ssgl{L}$. Using \eqref{Eq:Wignerd-risbo}, the FT in \eqref{Eq:GL-quadrature-fomrulation} can be expressed~(after rearranging the terms) as
\begin{equation}\label{Eq:GL-fomrulation_algo}
\sohc{f}{\ell}{m}{n} =  i^{n-m} \frac{2\ell +1}{2} \sum_{m'=-\ell}^{\ell}  \Delta_{m',m}^{\ell} \Delta_{m',n}^{\ell} \\ C_{m,n,m'},
\end{equation}
where
$C_{m,n,m'} = \sum_{v=0}^{L-1} A_{m,n}(\vartheta_v) q(\vartheta_v)   e^{im'\vartheta_v}$. For the computation of the FT, $A_{m,n}(\vartheta_v)$ is first be computed using \eqref{Eq:Amn_discrete} for all $|m|,|n|<L$ with the computational complexity $O(L^2\log L)$ for each $\vartheta_v$ and with the complexity $O(L^3\log L)$ and for all $\vartheta_v$. Once $A_{m,n}(\vartheta_v)$ is computed, $C_{m,n,m'} $ is then computed for all $|m|,|n|,|m'| <L$ in $O(L^4)$, which is then used to compute $\sohc{f}{\ell}{m}{n}$ using \eqref{Eq:GL-fomrulation_algo}, again with the complexity $O(L^4)$. Thus the overall complexity of FT is $O(L^4)$. For the proposed sampling scheme, we note that a fast algorithm of complexity $O(L^4)$ can also be developed by using the separation of variables approach as adopted in \cite{Kostelec:2008}.

\subsection{Inverse Fourier Transform~(IFT) --- Fast Algorithm}

For the computation of IFT, we rewrite \eqref{Eq:Amn}, using \eqref{Eq:Wignerd-risbo}, as
\begin{align}\label{Eq:Amn_IFT}
A_{m,n}(\vartheta_v) &= i^{n-m} \sum_{m'=-(L-1)}^{L-1} e^{im'\vartheta_v}{G_{m,n,m'}}.
\end{align}
where ${G_{m,n,m'}}={\sum_{\ell=0}^{L-1} \sohc{f}{\ell}{m}{n} \Delta_{m',m}^{\ell}\, \Delta_{m',n}^{\ell}}$, which does not depend on $\vartheta_v$, and therefore, can be computed for each
 $|m|,|n|,|m'|<L$ with the complexity $O(L^4)$. Using ${G_{m,n,m'}}$, $A_{m,n}(\vartheta_v)$ can be computed for all $|m|,|n|<L$  and for all sampling points $\vartheta_v$ using \eqref{Eq:Amn_IFT} in $O(L^4)$. Once $A_{m,n}(\vartheta_v)$ is computed, the signal $f(\varphi_u,\vartheta_v,\omega_w)$, sampled over either of the proposed sampling scheme, can be obtained with the complexity $O(L^3\log L)$ by employing FFT along $m$ and $n$ for each $\vartheta_v$, as given in \eqref{Eq:form-synthesis}. The overall complexity of the resulting algorithm to compute IFT is therefore $O(L^4)$.

\subsection{Computation of Wigner-$d$ functions}

In the computation FT and IFT, we need to address the computation of the fixed-angle Wigner-$d$ functions $\Delta_{m,n}^{\ell} = d^{\ell}_{m,n}(\pi/2)$, which are required to be computed for all $\ell<L$, and all $|m|,\,|n|\le \ell$. Let $\Delta^{\ell}$ denotes a matrix of size $(2\ell+1)\times (2\ell+1)$ with entries $\Delta_{m,n}^{\ell}$ for $|m|,|n|\le \ell$, which is computed for each $\ell=0,\,1,\,\dotsc,\,L-1$ using the recursion relation proposed in \cite{Trapani:2006} that recursively computes $\Delta^{\ell}$ from $\Delta^{\ell-1}$ with the computational complexity of $O(L^2)$. Since $\ell$ is of order $L$, the total complexity to compute $\Delta_{m,n}^{\ell}$  for all $0\le\ell<L$ and $|m|,|n|\le \ell$ is $O(L^3)$. In the computation of both FT and IFT, this recursion is useful as $\Delta_\ell$ matrices are computed recursively for $\ell=0,\,1,\,\dotsc,\,L-1$. In order to speed up the computation, we also employ the symmetry relations of Wigner-$d$ functions~\cite{Kennedy-book:2013}, which facilitate us in a way that we are only required to compute $(\ell+1)^2$ entries of the matrix $\Delta^{\ell}$. The remaining of $(2\ell+1)^2$ entries can be inferred using the symmetry relations.

\section{Accuracy Analysis}
\label{sec:analysis}

In order to analyse the numerical accuracy~(and stability) of the propose sampling scheme and the associated Fourier transforms, we carry out a following experiment: 1) obtain a band-limited test signal, denoted by $f_{\mathrm{t}}\in\lsphL{L}$, by generating its Fourier coefficients $\sohc{f_{\mathrm{t}}}{\ell}{m}{n}$ for $0<\ell<L, |m|,|n|\leq\ell$ with real and imaginary parts uniformly distributed in the interval $[-1,\,1]$, 2) apply the IFT to obtain the band-limited test signal $f_{\rm t}\in\lsphL{L} $ over the samples in the proposed scheme $\ssgl{L}$), 3) using the samples of the test signal, the FT is then computed which yields the reconstructed Fourier coefficients, denoted by $\sohc{f_{\mathrm{r}}}{\ell}{m}{n}$, 4) record the maximum reconstruction error $E_\mathrm{max}$ and the mean reconstruction error $E_\mathrm{mean}$, given by
\begin{align}
\label{Eq:exp1:errors:max}
E_\mathrm{max}
	&\dfn \max_{\ell,m,n}\big|\sohc{f_\mathrm{t}}{\ell}{m}{n} -
		\sohc{f_\mathrm{r}}{\ell}{m}{n}\big|, \\
E_\mathrm{mean}
	&\dfn \frac{1}{d_L} \sum_{\ell=0}^{L-1} \sum_{m=-\ell}^\ell \sum_{n=-\ell}^\ell \big|
		\sohc{f_\mathrm{t}}{\ell}{m}{n} - \sohc{f_\mathrm{r}}{\ell}{m}{n}\big| \label{Eq:exp1:errors:mean}.
\end{align}
The implementation the FT and IFT is carried out in double precision arithmetic in \text{\matlab}. For a particular band-limit in the range $2\leq L\leq 128$, we repeat the experiment for $5$ different test signals and obtain the average values of $E_\mathrm{max}$ and $E_\mathrm{mean}$, which are plotted in \figref{fig:accuracy}, that demonstrates that the IFT of any band-limited signal followed by the FT yields the same band-limited signal with errors on the order of the numerical precision. Hence, the proposed scheme supports exact computation of Fourier transforms on the rotation group.

\ifCLASSOPTIONonecolumn
\begin{figure}[t]
\centering
\includegraphics[width=0.65\textwidth]{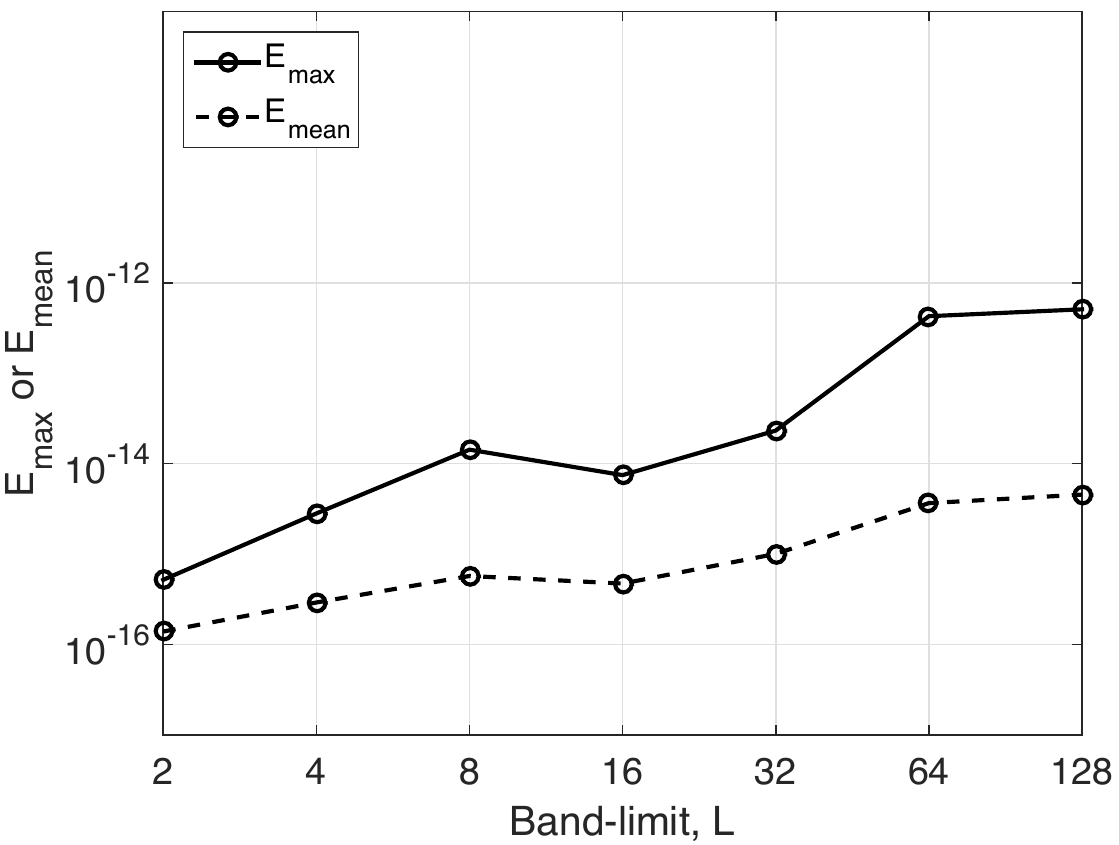}
\caption{Accuracy analysis: plots of the maximum error $E_\mathrm{max}$ and the mean error $E_\mathrm{mean}$, respectively given in \eqref{Eq:exp1:errors:max} and \eqref{Eq:exp1:errors:mean}, for band-limits in the range $2\leq L\leq 128$.}
\vspace{-2mm}
\label{fig:accuracy}
\end{figure}
\else
\begin{figure}[t]
\centering
\includegraphics[width=0.43\textwidth]{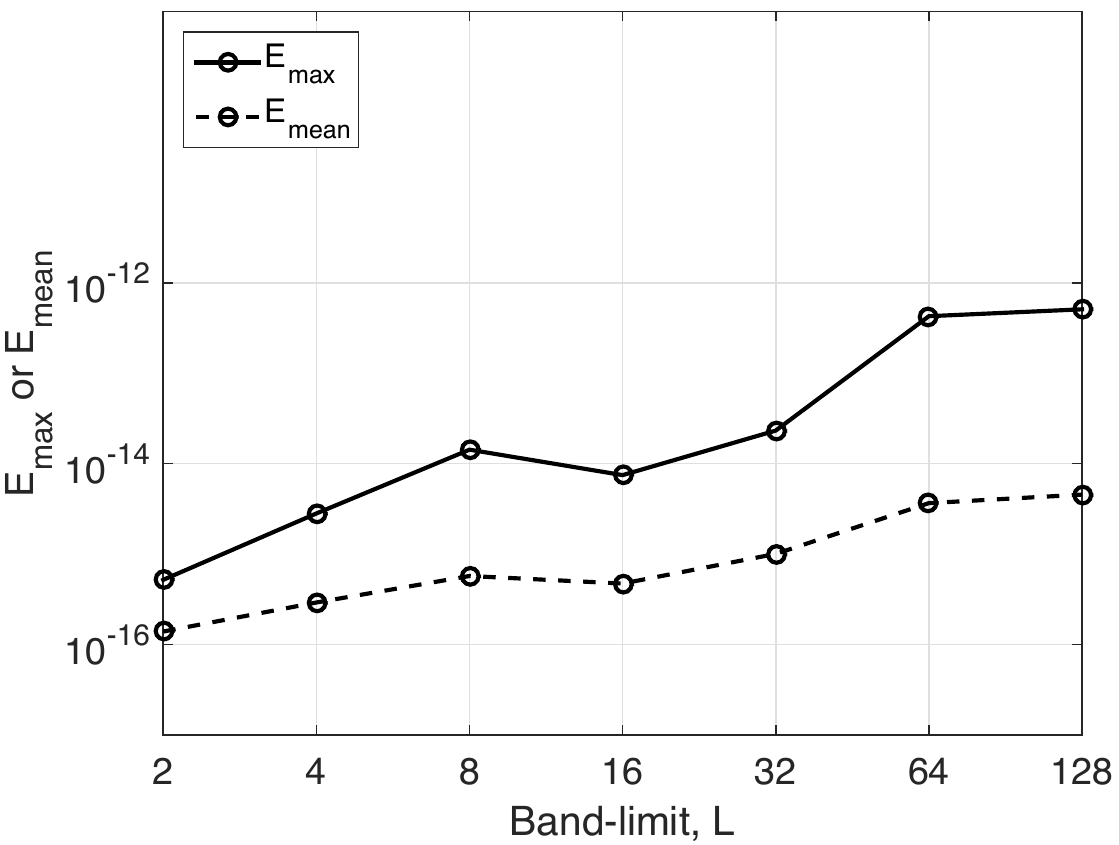}
\caption{Accuracy analysis: plots of the maximum error $E_\mathrm{max}$ and the mean error $E_\mathrm{mean}$, respectively given in \eqref{Eq:exp1:errors:max} and \eqref{Eq:exp1:errors:mean}, for band-limits in the range $2\leq L\leq 128$.}
\vspace{-2mm}
\label{fig:accuracy}
\end{figure}
\fi

\section{Conclusions}
\label{sec:conclusions}

We have developed a Gauss-Legendre quadrature based sampling scheme on the rotation group for the discretisation of band-limited signal defined on the rotation group such that the FT can be exactly computed from its samples. The proposed GL sampling scheme is asymptotically as efficient as the most efficient scheme developed recently. For the fast implementation of the FT and IFT, we have also developed fast algorithms of complexity $O(L^4)$. We have analysed the computation time and demonstrated the accuracy of the FT for the proposed sampling scheme up to band-limit $L=128$.


\end{document}